\documentclass[english]{revtex4}
\usepackage[T1]{fontenc}
\usepackage[latin9]{inputenc}
\usepackage{babel}

\begin{document}

\title{Dispersion relation of an electromagnetic wave in unmagnetized cold
plasma}

\author{Kushal Shah}

\email{atmabodha@gmail.com}

\address{Department of Electrical Engineering, IIT Madras, Chennai - 600036.}

\begin{abstract}
The dispersion relation of an electromagnetic wave in an unmagnetized
neutral plasma is well known to be $\omega^{2}=\omega_{p}^{2}+c^{2}k^{2}$.
A modified dispersion relation is presented taking into account the
ion restoring force in the transverse direction.
\end{abstract}
\maketitle

\section{Introduction}

It is well known that the dispersion relation of a plane electromagnetic
wave in an unmagnetized neutral cold plasma is given by $\omega^{2}=\omega_{p}^{2}+c^{2}k^{2}$.
To derive this, it has been assumed that the electron oscillations
in the transverse direction are of the same amplitude for all the
electrons on a plane perpendicular to a given point on the axis. Though
this assumption is true in theory, but questionable in practice. In
any real experiment, the electrons will oscillate with varying amplitude
in the transverse direction. This could happen due to boundary effects
or many other reasons. This effect can be small or large depending
on the problem at hand. For large amplitude variations, the above
dispersion relation will certainly breakdown. The question at hand
is: Does the above dispersion relation hold in the case of arbitrarily
small differences in the amplitude of single electron oscillations
in the transverse direction? Before we answer this, let us consider
the nature of langmuir oscillations.

The nature of langmuir oscillations is very peculiar. For a 1d neutral
cold plasma, the dispersion relation for langmuir waves is $\omega=\omega_{p}$.
This dispersion relation does not seem to depend on $k$, but there
is indeed a dependence. The correct way of writing the dispersion
relation for a langmuir wave is:\begin{eqnarray*}
\omega & = & 0\qquad k=0\\
 &  & \omega_{p}\qquad k\ne0\end{eqnarray*}
This is a very peculiar dispersion relation because $\omega$ is not
a smooth function of $k$. There is an abrupt jump at $k=0$. This
dispersion relation also holds in the nonlinear limit. The only thing
to make sure is that the wave amplitude should not be large enough
to cause wave-breaking.

As discussed before, when a plane electromagnetic wave propagates
in a plasma, we derive its dispersion relation assuming that there
is no variation in the transverse direction. In any practical system,
this is certainly not true. The assumption made is valid for systems
where the scale length of transverse variation is much larger compared
to the scale length of longitudinal variation. But no matter how small
the transverse variation may be, it will always be non-zero. And the
derivation of langmuir waves shows that all that we need to have electrostatic
oscillations at $\omega=\omega_{p}$ is a non-zero $k$. This $k$
can be as small as we like, but as long as it is $>0$, we will certainly
have electrostatic oscillations. This may sound trivial, but has very
important implications.

Thus, when an electromagnetic wave travels in a plasma, there will
always be electrostatic oscillations along the transverse direction.
And, as we will see, these electrostatic oscillations do modify the
dispersion relation of the electromagnetic wave. 

We first derive the known dispersion relation for em waves in a plasma.
Then, we add the electrostatic restoring force and show how it modifies
the em wave dispersion relation.

\section{Usual Derivation}

This derivation is given in any book on plasma physics%
\cite{nicholson}. I am reproducing it here for the benefit of the reader. 

The momentum equation is:\[
mn\left[\frac{\partial\vec{v}}{\partial t}+\vec{v}\cdot\vec{\nabla}\vec{v}\right]=qn\left[\vec{E}+\vec{v}\times\vec{B}\right]\]
And the two Maxwell's equations required are:\begin{eqnarray*}
\vec{\nabla}\times\vec{B} & = & \mu_{0}\vec{J}+\mu_{0}\epsilon_{0}\frac{\partial\vec{E}}{\partial t}\\
\vec{\nabla}\times\vec{E} & = & -\frac{\partial\vec{B}}{\partial t}\end{eqnarray*}
Under the linear approximation, we drop the nonlinear terms in the
momentum equation. Then the momentum equation becomes,\[
\frac{\partial\vec{v}}{\partial t}=\frac{q}{m}\vec{E}\]
\[
\Rightarrow\frac{\partial\vec{J}}{\partial t}=\frac{q^{2}n}{m}\vec{E}\]
Differentiating the first of the Maxwell's equation listed w.r.t.
time and substituting appropriately, we get,\[
-\vec{\nabla}\times\left(\vec{\nabla}\times\vec{E}\right)=\frac{q^{2}n}{m}\vec{E}+\frac{1}{c^{2}}\frac{\partial^{2}\vec{E}}{\partial t^{2}}\]
Now, since $\vec{\nabla}\cdot\vec{E}=0$, we get,\[
\vec{\nabla}^{2}\vec{E}-\frac{1}{c^{2}}\frac{\partial^{2}\vec{E}}{\partial t^{2}}=\frac{w_{p}^{2}}{c^{2}}\vec{E}\]
Now, on fourier transforming in space and time, we get the usual dispersion
relation:\[
\omega^{2}=\omega_{p}^{2}+c^{2}k^{2}\]

\section{A twist in the tale}

As explained in the introduction, as the electrons oscillate perpendicular
to the direction of propagation of the wave, they will experience
a restoring force due to ions due to infinitesimally small differences
in the amplitude of electron motion in the transverse direction. And
as long as wave breaking does not happen, this restoring force will
be proportional to the displacement of the particle. This is true
even in the nonlinear regime, as was shown by Dawson %
\cite{dawsom}. 

We assume that the em wave is polarized along the $\hat{x}$ direction
and travels along the $\hat{z}$ direction. Thus, the correct momentum
equation for the electrons should be,\[
\frac{\partial^{2}x}{\partial t^{2}}+w_{p}^{2}x=\frac{q}{m}\vec{E}=\frac{q}{m}E_{0}\cos\left(\omega t-kz\right)\]
\[
\Rightarrow x=-\frac{q}{m}\frac{E_{0}}{w^{2}-w_{p}^{2}}\cos\left(\omega t-kz\right)\]
\[
\Rightarrow qnv=\frac{q^{2}n}{m}\frac{wE_{0}}{w^{2}-w_{p}^{2}}\sin\left(\omega t-kz\right)\]
\[
\Rightarrow\frac{\partial J}{\partial t}=\frac{q^{2}n}{m}\frac{w^{2}E_{0}}{w^{2}-w_{p}^{2}}\cos\left(\omega t-kz\right)\]
The Maxwell's equation was,\[
-\vec{\nabla}\times\left(\vec{\nabla}\times\vec{E}\right)=\mu_{0}\frac{\partial\vec{J}}{\partial t}+\frac{1}{c^{2}}\frac{\partial^{2}\vec{E}}{\partial t^{2}}\]
\[
\Rightarrow\omega^{2}=\frac{\omega_{p}^{2}\omega^{2}}{\omega^{2}-\omega_{p}^{2}}+c^{2}k^{2}\]
If $\omega\gg\omega_{p}$, then the above expression reduces to the
conventional dispersion relation, $\omega^{2}=\omega_{p}^{2}+c^{2}k^{2}$. 

The above expression gives,\[
\omega^{4}-2\omega_{p}^{2}\omega^{2}-c^{2}k^{2}\omega^{2}+c^{2}k^{2}\omega_{p}^{2}=0\]
\begin{eqnarray*}
\Rightarrow\omega^{2} & = & \frac{\left(2\omega_{p}^{2}+c^{2}k^{2}\right)\pm\sqrt{\left(2\omega_{p}^{2}+c^{2}k^{2}\right)^{2}-4c^{2}k^{2}\omega_{p}^{2}}}{2}\\
 & = & \frac{\left(2\omega_{p}^{2}+c^{2}k^{2}\right)\pm\sqrt{4\omega_{p}^{4}+c^{4}k^{4}}}{2}\\
 & = & \left(\omega_{p}^{2}+\frac{c^{2}k^{2}}{2}\right)\pm\sqrt{\omega_{p}^{4}+\frac{c^{4}k^{4}}{4}}\end{eqnarray*}
For $ck\ll\omega_{p}$, we have,\begin{eqnarray*}
\omega^{2} & = & \left(\omega_{p}^{2}+\frac{c^{2}k^{2}}{2}\right)\pm\omega_{p}^{2}\sqrt{1+\frac{c^{4}k^{4}}{4\omega_{p}^{4}}}\\
 & \approx & \left(\omega_{p}^{2}+\frac{c^{2}k^{2}}{2}\right)\pm\omega_{p}^{2}\left(1+\frac{c^{4}k^{4}}{8\omega_{p}^{4}}\right)\end{eqnarray*}
If $ck\gg\omega_{p}$, we have,\begin{eqnarray*}
\omega^{2} & = & \left(\omega_{p}^{2}+\frac{c^{2}k^{2}}{2}\right)\pm\frac{c^{2}k^{2}}{2}\sqrt{1+\frac{4\omega_{p}^{4}}{c^{4}k^{4}}}\\
 & \approx & \left(\omega_{p}^{2}+\frac{c^{2}k^{2}}{2}\right)\pm\frac{c^{2}k^{2}}{2}\left(1+\frac{2\omega_{p}^{4}}{c^{4}k^{4}}\right)\end{eqnarray*}
As can be seem from the above expressions, $\omega<\omega_{p}$ is
an allowed solution of the dispersion relation. We can get this by
taking the -ve sign in the above expressions. This seems to an anomaly
since it is conventionally believed that for $\omega<\omega_{P}$
, the em wave should not be able to penetrate in the plasma. There
are two possibilities: Either the em wave with $\omega<\omega_{P}$
does actually penetrate into the plasma or there is some reason for
us to discard the -ve sign in the above expression. This question
can be answered only by more careful experiments in the future. 

One could argue saying that the above field is not purely electromagnetic
since we have also included an electrostatic component in the transverse
direction. It does not really matter by what name we call it. All
that matters is that the dispersion relation of a {}``initially''
transverse wave in a plasma seems to be given by \[
\omega^{2}=\frac{\omega_{p}^{2}\omega^{2}}{\omega^{2}-\omega_{p}^{2}}+c^{2}k^{2}\]

I would be glad to receive comments and views on this article through
email. I am open to both sides of the arguments. If you can find a
flaw in the above arguments and prove that this wrong, you are welcome.
And if you have evidence or a strong reason to say that the above
could be right, then of course, you are most welcome.

\end{document}